\documentclass[12pt,article]{revtex4}

\usepackage{epsfig}
\usepackage{bbm}
\usepackage{amssymb}
\usepackage{amsmath}

\newcommand{\bea}{\begin{eqnarray}}
\newcommand{\eea}{\end{eqnarray}}
\newcommand{\nn}{\nonumber}

\begin{document}
\rightline{HD-THEP-04-48}

\title{Axial Currents from CKM Matrix CP Violation
\footnote{Contribution to the Proceedings of Strong and Electroweak Matter 2004
(SEWM2004), Helsinki, Finland, June 16-19, 2004.}
}

\author{Thomas Konstandin}

\affiliation{Institut f\"ur Theoretische Physik,\\ 
Philosophenweg 16,\\
69120 Heidelberg, Germany\\
E-mail: T.Konstandin@ThPhys.Uni-Heidelberg.de}  

\maketitle

\centerline{\bf Abstract}

  The first principle derivation of kinetic transport equations suggests that 
a CP-violating mass term during the electroweak
phase transition can induce axial vector
currents. Since the important terms are of first order in gradients
there is a possibility to construct new rephasing invariants that 
are proportional to the CP phase in the
Cabibbo-Kobayashi-Maskawa matrix and to circumvent the 
upper bound of CP-violating contributions in the Standard Model, the 
Jarlskog invariant. 

%
%

\section{Introduction}

 All models that intend to describe the baryon asymmetry of the universe (BAU) 
by electroweak baryogenesis (EWB)\cite{Kuzmin:1985mm} depend on extensions of the Standard 
Model (SM) since the SM fails on the following grounds:

{\bf A)} First order phase transition:
Sakharov\cite{Sakharov:1967dj} pointed out that baryogenesis 
necessarily requires non-equilibrium 
physics. The expansion of the universe is too slow at the electroweak scale and 
one needs bubble nucleation during a first order EWPT. The phase diagram of the 
Standard Model is studied in detail\cite{Kajantie:1996qd},
and it is well known that there is no 
first order phase transition in the Standard Model for the experimentally 
allowed Higgs mass.

{\bf B)} Sphaleron bound:
To avoid washout after the phase transition, the {\it vev} of the broken Higgs field 
has to meet the criterion $\langle\Phi\rangle\gtrsim T_c$,
i.e. a strong first order phase transition. This results in the Shaposhnikov 
bound on the Higgs mass\cite{Farrar:hn}.

{\bf C)} Lack of CP violation:
Since the only source of CP violation in the Standard Model is the 
Cabibbo-Kobayashi-Maskawa (CKM) matrix (apart from the neutrino mass matrix, 
which provides an 
even tinier source of CP violation) one has to face that it is too weak to account 
for the observed magnitude of BAU.

In the following we will address the last point, the lack of sufficient 
CP violation, in the framework of kinetic theory. 

 The strong first order phase transition is assumed to occur at about $T_c \simeq 100$ GeV
and is parametrised by the velocity of the phase boundary (wall velocity) 
$v_w$ and its thickness $l_w$.

In this article we will focus on the following main points:

\begin{itemize}

\item We will demonstrate how CP violating sources can arise in semiclassical 
Boltzmann type equations.
\item We argue that the Jarlskog determinant as an upper bound of CP violation in the SM 
is not valid during the EWPT. 
\item We give a rough estimate of the CP violating sources during the EWPT and 
conclude that the source is by orders larger than considering the Jarlskog 
determinant but still insufficient to explain the magnitude of the BAU. 

\end{itemize}

\section{Axial Currents in Kinetic Theory}

   Starting point in kinetic theory are the exact Schwinger-Dyson equation 
for the two point functions in the closed time-path (CTP) formalism.
\begin{eqnarray}
&& \hskip -1.0 cm e^{-i\Diamond} \{ S_0^{-1} - \Sigma_R , S^< \} 
-  e^{-i\Diamond} \{ \Sigma^< , S_R \} \nonumber\\
&=& \frac{1}{ 2} e^{-i\Diamond} \{ \Sigma^< , S^> \} 
- \frac{1}{ 2} e^{-i\Diamond} \{ \Sigma^> , S^< \}, \\ 
&& \hskip -1.0 cm e^{-i\Diamond} \{ S_0^{-1} - \Sigma_R , {\cal A} \} 
-  e^{-i\Diamond} \{ \Sigma_A , S_R \} \nonumber 
=0, 
\end{eqnarray}
where we have used the definitions and relations
\begin{eqnarray}
 && S^{\bar t}:=S^{--}, \, S^t:=S^{++}, \, S^<:=S^{+-}, \, S^>:=S^{-+}, \nonumber \\
&& {\cal A}:=\frac{i}{2}(S^<-S^>),   
 S_R:=S^t-\frac{i}{2}(S^<+S^>),  \nonumber \\
&& 2\Diamond\{A,B\} := \partial_{X^\mu} A \partial_{k_\mu} B
-\partial_{k_\mu} A \partial_{X^\mu} B.
\end{eqnarray}
S denotes the Wightman function, $S_0^{-1}$ the free inverse propagator 
and $\Sigma$ the selfenergy. $S^{\pm\pm}$ denotes 
the entries of the $2\times 2$ Keldysh matrices and all functions depend 
on the average coordinate X and the momentum k.
  To simplify the equations, one can perform a gradient expansion. 
The terms on the left hand side will be expanded up to first order, 
whereas the collision terms on the right hand side vanish in equilibrium 
and are just kept up to zeroth order. The expansion parameter is formally 
$\partial_X/k$, which close to equilibrium and for typical thermal excitations 
reduces to $( l_w * T_c )^{-1}$.
 
We will not solve the full transport equations, but only look 
for the appearance of CP-violating source terms.

To start with, we consider a toy model with only one flavour and a  
mass term that contains a space dependent complex phase\cite{Kainulainen:2001cn}. 
The inverse propagator in a convenient coordinate system  reads
\bea
k_0 \gamma_0 + k_3 \gamma_3 + m_R(X_3) + i m_I(X_3) \gamma_5.
\eea

  Using the spin projection operator 
$P_s = \frac{1}{2}(1 + s\gamma_0\gamma_3\gamma_5)$
the Schwinger-Dyson equations can be decoupled and finally yield 
($m e^{i \theta} = m_R + i \, m_I$)
\bea
\left( k_0^2 - k_3^2 - m^2 + s \frac{ m^2\theta^\prime}{k_0} \right) Tr(\gamma_0 S^{<}_s) &=& 0 \\
\left( k_3 \partial_{X_3} - \frac{(m^2)^\prime}{2} \partial_{k_3} 
+s\frac{ (m^2\theta^\prime)^\prime }{2 k_0} \partial_{k_3} \right) Tr(\gamma_0 S^{<}_s) &=& Coll.
\eea

We see, that in our approximation, the quasi-particle picture is still valid, 
since the Wightman function fulfills an algebraic constraint. Furthermore, 
the first order corrections lead to some source term, that is  proportional 
to the complex phase of the mass and therefore CP violating.  

Performing the calculation with several flavours, one finds the 
generalization of this CP violating term, reading
$Tr(m^{\dagger\prime} m - m^\dagger m^\prime)$.

\section{Enhancement of CP Violation in the SM}

Jarlskog proposed an upper bound for CP violating effects 
in the Standard Model. Following her argument of rephasing invariants, 
the first CP violating quantity constructed out of the Yukawa couplings 
is the Jarlskog determinant\cite{Jarlskog:1988ii}
\bea
&& \hskip -0.3in Im \, det[\tilde m_u \,\tilde m_u^\dagger,
\tilde m_d \,\tilde m_d^\dagger]  \nn \\
&=& Tr(C m_u^4 C^\dagger m_d^4 C m_u^2 C^\dagger \tilde m_d^2) \nn \\
&\approx& -2J \cdot m_t^4 m_b^4 m_c^2 m_s^2, \label{jarlsdet}
\eea
When applied to the case of electroweak baryogenesis, one finds 
the upper bound of the BAU\cite{Farrar:sp}
\bea
&[\frac{g_W^2}{2M_W^2}]^7 J \cdot m_t^6 m_b^4 m_c^2 m_s^2 
\approx 10^{-22},& \label{shapbound}
\eea
Though, two assumptions, that are  needed  for this argument are 
not fulfilled  during the electroweak phase transition\cite{Konstandin:2003dx}.

{\bf A)} Since the mass matrix is space dependent, one needs space 
dependent diagonalization matrices to transform to the mass eigenbasis. This
leads to new physical relevant quantities, that can as well be CP violating. 
As a generalization of the CP violating source term in the kinetic
toy model above, we found $Tr(m^{\dagger\prime} m - m^\dagger m^\prime)$.
However in the Standard model this term vanishes at tree level, for 
the mass matrix is proportional to its derivative.

{\bf B)} The argument of Jarlskog is based on the fact, that 
the examined quantity is perturbative in the Yukawa coupling. 
The calculation of the selfenergy in a thermal plasma 
involves integrations over divergent logarithms, of the form   
\bea
\hskip -0.2in h_2(\omega,\kappa) 
&=& \frac{1}{\kappa} \int_0^\infty \frac{d|{\bf p}|}{2\pi}
 \Big( \frac{|{\bf p}|}{\epsilon_h} L_2(\epsilon_h,|{\bf p}|) f_B(\epsilon_h) \nn \\
&& \hskip 0.5in - \frac{|{\bf p}|}{\epsilon_u} L_1(\epsilon_u,|{\bf p}|) f_F(\epsilon_u) \Big) . \nn
\eea 
\bea
\hskip -0.3in L_{1/2}(\epsilon,|{\bf p}|) 
&=& \log \left( \frac
{\omega^2 - \kappa^2 \pm \Delta + 2 \epsilon \omega + 2 \kappa |{\bf p}|}
{\omega^2 - \kappa^2 \pm \Delta + 2 \epsilon \omega - 2 \kappa |{\bf p}|} \right) \nn \\
&& + \log \left( \frac
{\omega^2 - \kappa^2 \pm \Delta - 2 \epsilon \omega + 2 \kappa |{\bf p}|}
{\omega^2 - \kappa^2 \pm \Delta - 2 \epsilon \omega - 2 \kappa |{\bf p}|}
\right), \nn
\eea
that lead to a significant space dependence of the selfenergy.
\begin{figure}[ht]
\centerline{\epsfxsize=3.1in\epsfbox{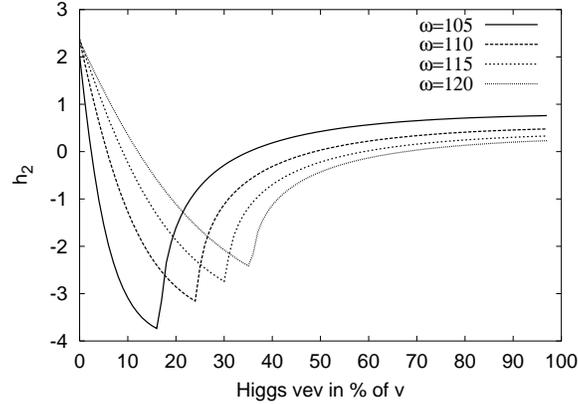}}   
\caption{Dependence of $h_2$ on the Higgs vev $\Phi$ in \% 
of its value $\Phi^0=246$ GeV at T=0. The external energies and momenta 
are fixed at $\omega=105$ GeV to  $\omega=120$ GeV, k=100 GeV, 
the mass of the quark in the loop is $m_u=100$ GeV.
 \label{inter}}
\end{figure}
Since the space dependence is due to a resonance with the plasma 
particles, the selfenergy is highly sensitive to the quark
masses and the W mass, that both change continously in the wall profile.
\begin{figure}[ht]
\centerline{\epsfxsize=3.1 in\epsfbox{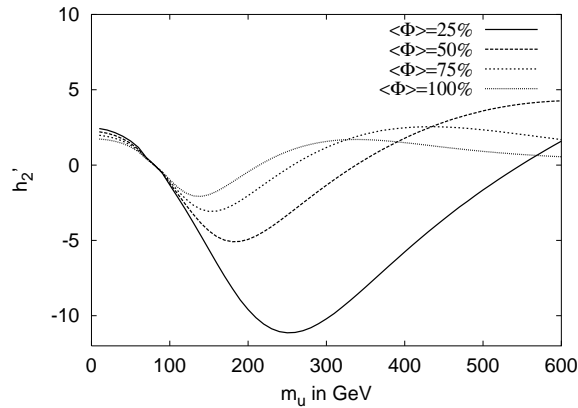}}   
\caption{Dependence of $h_2^\prime$ on the mass of the quark 
in the loop with an on-shell external quark of mass $m_e=4$ GeV. 
The Higgs vev is chosen in a range of 25\% to 100\% of its value 
in the broken phase at T=0. 
\label{inter2}}
\end{figure}
However since CP violating effects only appear as an interference of the two loop
and the one loop term, an estimation of the source term leads to  the upper bound\cite{Konstandin:2003dx}
$$ \frac{\delta\omega}{\omega} \sim J \cdot 
m_t^4 m_s^2 m_b^2 m_c^2 \frac{ \alpha_w^3 h_2^\prime}
{m_W^8 l_w T^3} \approx 10^{-15}.$$

We conclude, that the axial current is enhanced seven orders 
 in magnitude. Still the CP-violating source due to the CKM matrix might be too weak 
to account for the BAU.
\vskip 0.3 cm
{\bf Acknowledgements.}
I would like to thank M.G.~Schmidt and T.~Prokopec for the nice collaboration.

\end{document}